\documentclass{aip-cp}

\usepackage{graphicx}

\begin{document}

\title{On the Origin of the $XYZ$ Mesons}

\author[aff1]{A. Valcarce\corref{cor1}}
\author[aff2]{J.~Vijande}
\eaddress{javier.vijande@uv.es}

\affil[aff1]{Departamento de F\'\i sica Fundamental and IUFFyM, Universidad de Salamanca, E-37008
Salamanca, Spain.}
\affil[aff2]{Departamento de F\'{\i}sica At\'{o}mica, Molecular y Nuclear, Universidad de Valencia (UV)
and IFIC (UV-CSIC), Valencia, Spain.}
\corresp[cor1]{Corresponding author: valcarce@usal.es}

\maketitle

\begin{abstract}
In this talk we present a mechanism giving rise to exotic $XYZ$ four-quark states in the meson spectra within a 
constituent quark model approach. We discuss its generalization to five-quark states in the heavy baryon
sector. Finally, we revise some other works in the literature and experimental data where this mechanism may be working.
\end{abstract}

\section{INTRODUCTION}

Before the discovery of the $X(3872)$~\cite{Cho03}, the hadronic experimental 
data were classified either as $q\overline{q}$ or $qqq$ states according to 
$SU(3)$ irreducible  representations based on Gell-Mann conjecture~\cite{Gel64}. 
However, since 2003 more than twenty newly observed 
meson resonances reported by different experimental collaborations appeared, close to a two-meson threshold,
presenting properties that make a simple quark-antiquark structure unlikely~\cite{Bod13}, the so-called $XYZ$ states. 
Although this observation could be coincidental due to the large number of thresholds in the energy region
where the $XYZ$ mesons have been reported, it could also point to a close relation between some particular 
thresholds and resonances contributing to the standard quark-antiquark heavy meson spectroscopy. 

More than a decade has elapsed since the discovery of the X(3872), and no compelling
explanation for the pattern of $XYZ$ mesons has emerged.
Several alternatives have been proposed in the literature to address these $XYZ$ states~\cite{Bra13}, being the most 
common ones conventional quarkonium~\cite{Eic04}, which consists of a color-singlet heavy quark-antiquark
pair: $(Q\bar Q)_1$; meson-meson molecules~\cite{Tor94}, which consists of color-singlet $Q \bar q$ and $\bar Q q$ mesons 
bound by hadronic interactions: $(Q\bar q)_1 + (\bar Q q)_1$; quarkonium 
hybrids~\cite{Clo05}, which  consists of a color-octet $Q\bar Q$ pair to  which  a
gluonic excitation is bound: $(Q\bar Q)_8 + g$; four-quark states~\cite{Vij07a}, which 
consists of a $Q\bar Q$ pair and a light quark $q$ and antiquark $\bar q$ bound by 
interquark potentials into a color singlet: $(Q\bar Q q\bar q)_1$; and diquarkonium~\cite{Mai05},  
which consists of a color-antitriplet $Qq$ diquark and a color-triplet $\bar Q \bar q$ 
diquark bound by the QCD color force: $(Qq)_{\bar 3} + (Qq)_3$. 
However, none of the models that has been proposed provide a plausible
pattern for all the $XYZ$ mesons that have been observed.

In this talk we analyze heavy hadron spectroscopy beyond open flavor thresholds considering higher order 
Fock space components looking for a general pattern of the $XYZ$ states. We highlight the pivotal
role played by hadron-hadron thresholds in heavy hadron spectroscopy and under which conditions, if any, 
they could cause a resonance to appear. We will analyze the thresholds open for 
four-quark states contributing to heavy meson spectroscopy and we will show how 
these thresholds may entangle so a bound four-quark state or a resonance may emerge and 
the conditions required for that~\cite{Vij14}. We will extend our arguments to the heavy baryon spectra, 
where one could also find contributions with a involved structure,
such as compact five--quark states beyond simple $ND$ resonances~\cite{Vij09,Car14}.
The study of these contributions requires from a full coupled-channel approach including all
possible physical states contributing to a given set of quantum numbers $(T,J)$, as has
been demonstrated in Ref.~\cite{Car11} for the charmonium spectrum.

\section{FORMALISM}

Standard mesons ($q\bar q$) and baryons ($qqq$) are the only clusters of quarks where it is not possible to construct a 
color singlet using a subset of their constituents. This, however, is not the case for multiquark combinations, and in 
particular for four--quark and five--quark states addressing the meson and baryon spectra, respectively. Thus, when 
dealing with higher order Fock space contributions to
hadron spectroscopy, one has to discriminate between possible multiquark bound states or resonances 
and simple pieces of the hadron--hadron continuum. For this purpose, one has to analyze the two--hadron 
states that constitute the threshold for each set of quantum numbers.
These thresholds have to be determined assuming quantum number conservation within exactly the same 
scheme (parameters and interactions) used for the multiquark calculation. If other models, parametrizations 
or experimental masses are used, then multiquark states might be misidentified as members
of the hadron spectra while being simple pieces of the continuum. 

Given a general four--quark state, ($q_1q_2\bar q_3\bar q_4$), two different thresholds are 
allowed, $(q_1\bar q_3)(q_2\bar q_4)$ and $(q_1\bar q_4)(q_2\bar q_3)$. If the four--quark system 
contains identical quarks, like for instance $(QQ\bar n\bar n)$ (in the following $n$ 
stands for a light quark and $Q$ for a heavy $c$ or $b$ quark), the two thresholds are
identical, i.e., $(Q \bar n)(Q\bar n)$. The importance of this particular feature lies on the fact that 
a modification of the four--quark interaction would not necessarily translate into the 
mass of the two free-meson state. Therefore, the unique necessary condition required to have 
a four--quark bound state would be the existence of a sufficiently attractive interaction between 
quarks that do not coexist in the two free-meson states. This hypothesis was 
demonstrated by means of the Lippmann--Schwinger formalism in Ref.~\cite{Car11}, concluding the 
existence of a single stable isoscalar doubly charmed meson with quantum numbers $J^P=1^+$. 
    
For those cases containing a heavy quark and its corresponding heavy antiquark $(Q n \bar Q \bar n)$ the 
situation is remarkably different. Two different thresholds are allowed, namely $(Q\bar Q)(n\bar n)$ and $(Q\bar n)(n\bar Q)$. 
It has been proved~\cite{Ber80} that ground state solutions of the Schr\"odinger ($q_1\bar q_2$) two--body problem 
are concave in $(m_{q_1}^{-1}+m_{q_2}^{-1})$ and hence $M_{Q\bar n}+M_{\bar Q n}\geqslant M_{Q\bar Q}+M_{n\bar n}$. 
This property is enforced both by nature\footnote{$M_{D^*}+M_{\bar D^*}=4014$ MeV $\geqslant M_{J/\psi}+M_{\omega}=3879$ MeV} 
and by all models in the literature unless forced to do otherwise. 
Although this relation among ground-state masses makes the assumption of a strictly flavor
independent potential, one should bear in mind that ground states of heavy mesons are perfectly
reproduced by a Cornell-like potential~\cite{Clo03}, that it is flavor independent.
The color-spin dependence of the potential would go in favor of this relation
for ground states (spin zero) because the color-spin interaction is attractive
for spin zero and comes suppresed as $1/(m_im_j)$, making even lighter the mesons
on the right hand side. Regarding the spin independent part,
the binding of a coulombic system is proportional to the reduced mass of the
interacting particles. Thus, for a two-meson threshold with a heavy-light light-heavy 
quark structure, the binding of any of the two mesons is proportional to the reduced 
mass of each meson, being close
to the mass of the light quark. However, if the two-meson state presents a
heavy-heavy light-light quark structure, the binding of the heavy-heavy meson increases 
proportionally to the mass of the heavy particle while that of the light-light meson remains constant,
becoming this threshold lighter than the heavy-light light-heavy two-meson structure.
Thus, it implies that in all relevant 
cases the lowest two-meson threshold for any $(Q n \bar Q \bar n)$ state will be the one made of quarkonium-light 
mesons, i.e., $(Q\bar Q)(n\bar n)$ (see Fig. 1 of Ref.~\cite{Car12}).
The interaction between the heavy, $(Q\bar Q)$, and light, $(n\bar n)$, mesons forming the lowest threshold is 
almost negligible, due to the absence of a light pseudoscalar exchange mechanism between them~\cite{Tor94}. 
Hence, any attractive effect in the four--quark system must have its origin in the interaction of the higher channel $(Q\bar n)(n\bar Q)$
or due to the coupled channel effect of the two thresholds, $(Q\bar Q)(n\bar n)\leftrightarrow(Q\bar n)(n\bar Q)$~\cite{Lut05}.

Similar arguments could be used in the case of the heavy baryon spectra. 
Given a general five--quark state contributing to the heavy (charm or bottom) baryon spectrum, ($nnnQ\bar n$), two different thresholds are 
allowed, $(nnn)(Q\bar n)$ and $(nnQ)(n\bar n)$.
A straightforward generalization of the concave behavior in $(m_{q_1}^{-1}+m_{q_2}^{-1})$ of the ground state solutions 
of the Schr\"odinger ($q_1\bar q_2$) two--body problem 
to the five--quark system could be obtained within
a quark-diquark model if $m_{q_1} \le m_{q_2} \le m_{q_3}$. Then
$M_{q_3 \bar q_2} + M_{q_1 \bar q_1} \le M_{q_3 \bar q_1} + M_{q_1 \bar q_2}$,
because the intervals in $1/ \mu$ of the left hand side and right hand side
have the same middle, but the left hand side one is wider that the right hand side one.
Now, in a crude quark-diquark model, one can translate this as
$M_{q_3 q_1 q_1} + M_{q_1 \bar q_1} \le M_{q_3 \bar q_1} + M_{q_1 q_1 q_1}$, as it is 
observed in Fig. 5 of Ref.~\cite{Car14}, except for the higher spin states where the angular momentum 
coupling rules impose further restrictions~\footnote{We thank
to J.~M. Richard for this simple and nice argument that does not make any assumption on the shape of 
the interaction, linear or not, although it assumes a quark-diquark ansatz.}.

In those energy regions where bound and unbound solutions of the multiquark hamiltonian coexist, methods based on 
infinite expansions become inefficient to hunt a bound state close to an unbound solution, because too many basis 
states would be required to disentangle them. The most 
interesting case where this may happen is in the vicinity of a two--hadron threshold, because 
both the two free--hadron state and a feasible slightly bound multiquark state are solutions of the same 
hamiltonian. Such cases have been studied by means of the Lippmann-Schwinger equation in Ref.~\cite{Car09},
looking at the Fredholm determinant $D_F(E)$ at zero 
energy~\cite{Gar87}. If there are no interactions then $D_F(0)=1$, 
if the system is attractive then $D_F(0)<1$, and if a
bound state exists then $D_F(0)<0$. All states made of $S$ wave 
($Q\bar n$)--($n\bar Q$) mesons up to $J = 2$ were scrutinized. A few channels 
were found to be slightly attractive, 
$D\bar{D}$ with $(I)J^{PC}=(0)0^{++}$, $D\bar{D}^*$ with $(0)1^{++}$ 
and $D^*\bar{D}^*$ with $(0)0^{++}$, $(0)2^{++}$, and $(1)2^{++}$, close 
to the results of Ref.~\cite{Tor94}. However, the only bound state appeared in the 
$(I)J^{PC}=(0)1^{++}$ channel as a consequence of the coupling between 
$D\bar D^*$ and $J/\Psi \omega$ two--meson channels. 

The conclusions of Refs.~\cite{Vij07a} and~\cite{Car09} point to a convoluted four--quark 
molecular structure with a dominant $D\bar D^*$ component for the $X(3872)$. However, this is not the only supernumerary state that 
has appeared in the charm and bottom sectors during the last years. A comprehensive list of 
such $XYZ$ mesons and their properties can be found in Ref.~\cite{Bod13}. 
20 states have been reported by different experimental collaborations in the charmonium sector 
above the $D\bar D$ threshold, 15 of them neutral and 5 charged. In the bottom sector 2 charged 
and 1 neutral state have been reported. Of those 23 states only 8 have been observed independently 
by two different collaborations and with significance greater than 5$\sigma$: $X(3872)$, $X(3915)$, $\chi_{c2}(2P)$, $G(3900)$, $Y(4140)$, $Y(4260)$, 
$Y(4360)$, and $Z_c^+(3900)$ (see Table I of Ref.~\cite{Bod13}). Therefore, although some of them might not resist cross-check 
examination by independent experimental collaborations, others are clearly established as 
real resonances and therefore they have to be accounted for in any description of the meson spectra.
While some of them, like the $\chi_{c2}(2P)$, seem to fit nicely within a naive quark-antiquark scheme, others do not. 

\section{RESULTS AND DISCUSSION}

\begin{figure}[tb]
\centerline{\includegraphics[height=12cm]{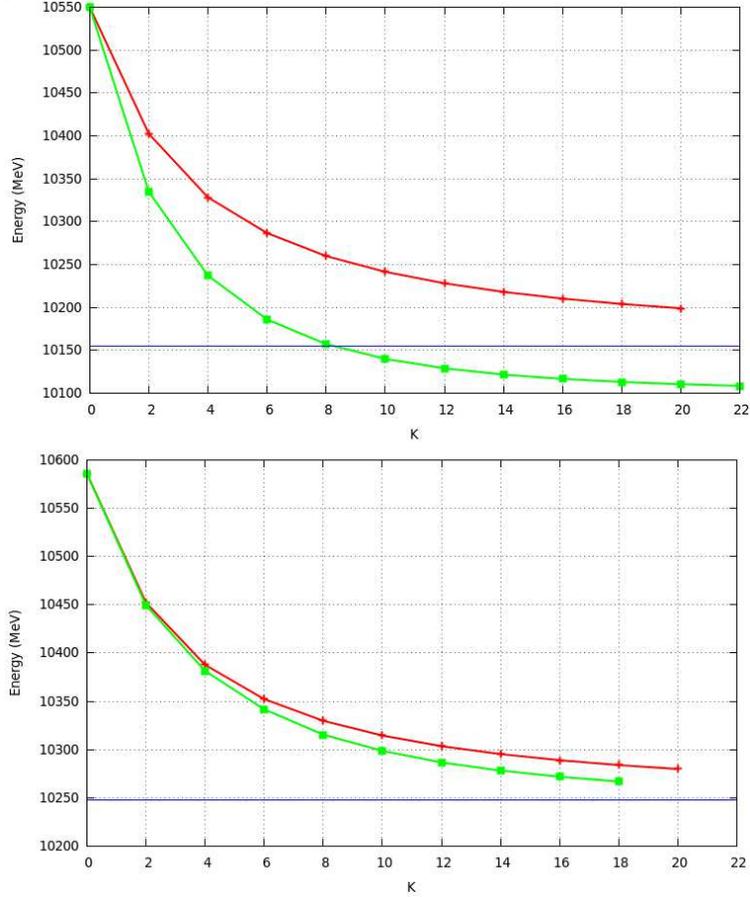}}
\caption{Convergence of $b\bar bn\bar n$ with quantum numbers $L=0$, $S=1$, $C=+1$ $P=+1$, $I=0$ (upper panel) 
and $I=1$ (lower panel). Red lines correspond to the case where the thresholds are non-degenerate and green 
lines to the case where they are almost degenerate.}
\label{fig1}
\end{figure}

When four-quark components are considered in the wave function of charmonium, there are 72 $c\bar c n\bar n$ 
combinations of quantum numbers for total orbital angular momentum $L<3$. Therefore, the question is not 
whether it is possible to design a model, or a formalism, able to match one of the newly observed $XYZ$ states 
with a particular set of these quantum numbers but to understand where the attraction comes from 
and to explain the systematic that predicts where, if anywhere, 
experimentalists and theoreticians alike should look into.
Since the lowest threshold interaction is rather weak, the possibility to obtain
bound states may only stem from the vicinity of an attractive $(c\bar n)(n\bar c)$ threshold 
coupled sufficiently as to bind the system as it occurs with the X(3872)~\cite{Car09}. 
Although for heavier mesons interactions are more attractive~\cite{Car12}, the effect of
channel coupling may not be enough to favor binding. Thus,
to check the efficiency of this mechanism, we solved with the HH formalism 
the bottom counterpart of the $X(3872)$, $b\bar bn\bar n$ with quantum numbers 
$L=0$, $S=1$, $I=0$, $C=+1$, and $P=+1$. Within the CQCM of Ref.~\cite{Vij05} the corresponding lowest 
thresholds, $B\bar B^*$ (10611 MeV) and $\Upsilon\omega$ (10155 MeV), are 456 MeV apart. We 
show in Fig.~\ref{fig1}(upper panel) the convergence pattern of the energy of the
four-quark system as a function of the 
hyperangular momenta $K$. It can be clearly seen how the energy of the four--quark system (red line)
is converging to the lowest threshold $\Upsilon\omega$ (horizontal blue line), what is a sharp signal
of an unbound state. One could however play around with the model parameters 
to almost degenerate both thresholds by adding 
attraction in the heavy-light $bn$ sector\footnote{We have slightly increased
the $\alpha_s(bn)$ strong coupling constant from 0.55 to 0.85, what would move the gap
between thresholds without changing the lowest threshold}, what would also increase the coupled channel effect
strengthening the $B\bar B^*\leftrightarrow \Upsilon\omega$ transition interaction. 
When this is done we note (green line) 
that the energy drops below threshold and a bound state emerges. One may wonder if only the close-to-degeneracy of the
thresholds is sufficient to bind this type of four--quark systems. If this would be the case, then the charged partner 
of this four--quark state ($I=1$) should behave exactly in the same manner. However, amazingly this is not so. 
In Fig.~\ref{fig1}(lower panel) we depict the convergence of the isovector state 
as a function of $K$ for both cases, non-degenerate thresholds (red line) 
and almost degenerate ones (green line). In this case the lowest threshold would be $\Upsilon\rho$ (10248 MeV). It can be observed that in both cases the 
four--quark state converges to the lowest threshold and does not form a bound state.

\begin{figure}[tb]
\centerline{\includegraphics[width=10cm]{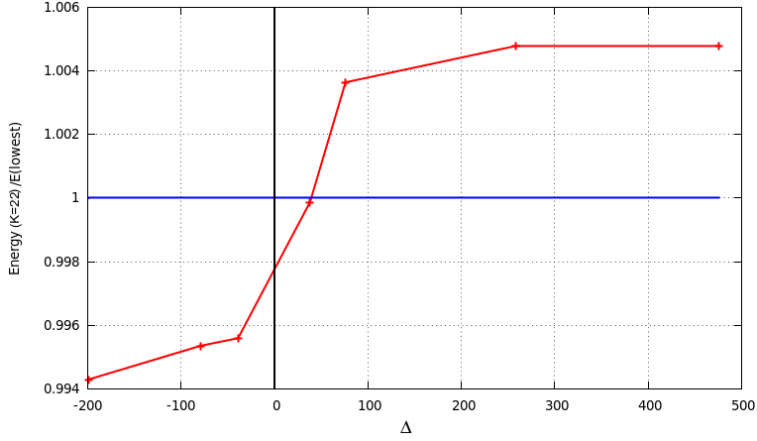}}
\vspace*{-0.5cm}
\caption{Energy divided by the lowest two-meson threshold of $b\bar bn\bar n$ with quantum 
numbers $L=0$, $S=1$, $C=+1$ $P=+1$, $I=0$ as a function of $\Delta = E[(b\bar n)(n\bar b)]-E[(b\bar b)(n\bar n)]$.}
\label{fig2}
\end{figure}

Thus, when the $(Q\bar Q)(n\bar n$) and $(Q\bar n)(n\bar Q)$ thresholds are sufficiently far away, which means that the interaction in the higher
$(Q\bar n)(n\bar Q)$ state is weak and therefore the coupled channel effect small, no bound states are found for any set of parameters. 
However, when the thresholds move closer, i.e., the attraction in the higher two-meson state
and the coupled channel strength are simultaneously increased, bound states may appear for a subset of 
quantum numbers. Hence, threshold vicinity is a required but not sufficient 
condition to bind a four--quark state. An additional condition is 
required to allow the emergence of such bound states. Such condition is the 
existence of an attractive interaction in the higher $(Q\bar n)(n\bar Q)$ 
two--meson system that would also give rise to a strong $(Q\bar Q)(n\bar n)\leftrightarrow(Q\bar n)(n\bar Q)$ coupling. 
To neatly illustrate this conclusion we return to the four--quark 
state $b\bar bn\bar n$ with quantum numbers $L=0$, $S=1$, $I=0$, 
$C=+1$, and $P=+1$. In Refs.~\cite{Car12,Car09} it was proved that the interaction provided 
by the CQCM model is attractive for these quantum numbers in the 
charm sector, although does not present a $D\bar D^*$ bound state. In the bottom sector the attraction is enhanced~\cite{Car12}. Thus, we have solved 
the four--body problem as a function of the threshold energy difference, 
$\Delta = E[(b\bar n)(n\bar b)]-E[(b\bar b)(n\bar n)]$, ranging from 500 MeV to $-$200 MeV. 
We show in Fig.~\ref{fig2} the energy normalized to the mass of the lowest two-meson threshold for each particular case, 
i.e., values smaller than 1 will point to a bound state and those larger to an unbound state. 
In this case, the results can be separated into two distinct categories: (i) $\Delta\gtrsim 50$ MeV 
and (ii) $\Delta\lesssim 50$ MeV. When the thresholds are separated by more than 50 MeV, 
the attractive interaction in the $B\bar B^*$ system and the coupled channel effect is not sufficient to overcome 
the threshold energy gap, and therefore the four--quark system evolves to an unbound two--meson state. 
However, when the thresholds get closer, even reversed for $\Delta<0$, the system 
becomes a compact four--quark state. Of particular interest are those cases 
where $\Delta\simeq50$ MeV in Fig.~\ref{fig2}. In this case the attraction in the higher channel together
with the coupled channel effect barely overcomes 
the threshold energy difference, and hence its wave function becomes strongly
entangled. This would generate a molecular state just close to threshold.

It should also be noted that $\Delta_0$, the $\Delta$ value for which Energy($4q$)/E(lowest) is equal to one, will change depending on 
the particular set of quantum numbers considered. However, $\Delta_0$ = 0 implies that no binding energy is provided by the upper threshold 
and the off-diagonal terms and therefore such solution will not correspond to a bound state. 
In that case our calculation will be simply providing a piece of the meson-meson unbound continuum.

The same argument has been drawn in Ref.~\cite{Bar15} where the {\it supercharmonium} states have 
been introduced, experimental resonances that appear to contain a $(c\bar c)$ pair, but have other 
properties that preclude a description in terms of only $(c\bar c)$ (idealized charmonium) basis states. Four-quark
supercharmonium configurations occur very near, or even below, the lowest S-wave threshold
for production of meson pairs. These states evade Coleman's argument in 1979~\cite{Col79}, who used the $1/N_c$
expansion of QCD to show that four--quark color singlets tend to propagate as pairs of mesons. One of these 
superchrmonium states could be the $Z(4475)$, that may appear as a linear superposition
of $D\bar D^*$ mesons with one of them in a radially excited state $2S$. In a simple
one-pion exchange model the $D \to D + \pi$ vertex is forbibben while the $D^* \to D + \pi$
is allowed. This gives rise to weak diagonal interactions but a strong mixing between
two thresholds, $D(1S)\bar D^*(2S)$, 4482 MeV/c$^2$, and $D^*(1S) \bar D (2S)$, 4433 MeV/c$^2$. 
Besides, the small mass difference between the
thresholds due to the diminishing of the hyperphine splitting when increasing the radial
excitation, leads to an amplification of the binding. The mechanism would be the same
working for the $X(3872)$ but with a complete degeneracy between $D(1S)\bar D^*(1S)$
and $D^*(1S) \bar D(1S)$.

A similar situation could be found in the case of the heavy baryon spectrum.
An important source of attraction might be the coupled-channel effect of the two 
thresholds, $(nnn)(Q\bar n)\leftrightarrow(nnQ)(n\bar n)$~\cite{Lut05}.
Thus, to check the efficiency of this mechanism, we have performed a calculation
considering all physical channels, $(nnn)(Q\bar n)$ and $(nnQ)(n\bar n)$.
When the $(nnn)(Q\bar n)$ and $(nnQ)(n\bar n)$ thresholds are sufficiently 
far away, the coupled-channel effect is small, and bound states are not found. 
However, when the thresholds move closer, the coupled-channel strength is increased, 
and bound states may appear for a subset of quantum numbers. Hence, threshold vicinity is 
again a required but not sufficient 
condition to bind a five--quark state. Under these conditions, there are the channels 
with high spin $J^P=5/2^-$ the only ones that may lodge a 
compact five-quark state for all isospins~\cite{Car14}. The reason stems on the reverse of the ordering of the
thresholds, being the lowest threshold $(nnn)(Q\bar n)$ the one with the more attractive
interaction. Of particular interest is the $(T)J^P=(2)5/2^-$ state, that survives
the consideration of the break apart thresholds. It may
correspond to the $\Theta_c(3250)$ pentaquark found by the QCD sum rule analysis of Ref.~\cite{Alb13}
when studying the unexplained structure with a mass of 3250 MeV/c$^2$ in the $\Sigma_c^{++} \pi^- \pi^-$ 
invariant mass reported recently by the BABAR Collaboration~\cite{Lee12}. 
Such state could therefore be a consequence of the close-to-degeneracy of the lowest thresholds with
$I=2$ and $J^P=5/2^-$,  $\Delta D^*$ and $\Sigma^*_c \rho$ and the attractive interaction
of the $\Delta D^*$ system~\cite{Car14}. It could be detected by the propagation of $D$ mesons in nuclear matter as
an $S$ wave $\Delta D^*$ system and it thus constitutes a challenge for the $\bar P$ANDA Collaboration.

\section{CONCLUSIONS}

To summarize, we have presented a plausible mechanism for the origin of the $XYZ$ mesons in the 
heavy meson spectra within a standard quark-model picture. Its generalization to the heavy baryon sector 
has been analyzed. The existence of open flavor 
two--hadron thresholds is a feature of the heavy hadron spectra that needs to be considered 
as a relevant ingredient into any description of the plethora of new states reported in 
heavy meson and baryon spectroscopy. They might be, at a first glance, identified 
with simple quark-antiquark or three-quark states, however in some cases their energies and decay 
properties do not match such oversimplified picture. 
Our results prove the relevance of higher order Fock space components
through the allowed two-hadron thresholds. On the one hand, 
one has the lower $(Q\bar Q)(n\bar n)$ and $(nnn)(Q\bar n)$ systems, made 
by almost noninteracting hadrons, that constitutes the natural breaking apart end-state. On the other hand, the 
higher $(Q\bar n)(n\bar Q)$ and $(nnQ)(n\bar n)$ systems appear. When there is an attractive interaction characterizing the 
upper systems combined with a strong enough coupling, together with the vicinity of the two allowed thresholds, 
a multiquark bound state may emerge.
As one can see the mechanism proposed is restrictive enough as not
predicting a proliferation of bound states when explaining the existence
of an hypothetical molecular structure. 
Once this is performed, the present
experimental effort with ongoing experiments at BESIII, current analyses by the LHC collaboration
and future experiments at Belle II and Panda together with the very impressive results that are 
being obtained by lattice gauge theory calculations~\cite{Liu12} may confirm the theoretical 
expectations of our quark-model calculation pattern that will provide with a deep understanding 
of low-energy realizations of QCD.

\section{ACKNOWLEDGMENTS}
This work has been partially funded by the Spanish Ministerio de Educaci\'on y Ciencia 
and EU FEDER under Contract No. FPA2013-47443,
by the Spanish Consolider-Ingenio 2010 Program CPAN (CSD2007-00042) and by Generalitat 
Valenciana Prometeo/2009/129. A.V. thanks finantial support from the Programa
Propio I of the University of Salamanca.

\end{document}